\documentclass[english,twocolumn,prl,amssymb,amsmath,superscriptaddress]{revtex4}
\usepackage[latin9]{inputenc}
\usepackage{graphicx,color}
\usepackage{hyperref}
\usepackage{babel}
\usepackage{booktabs}
\usepackage[normalem]{ulem}

\newcommand{\be}{\begin{equation}}
\newcommand{\ee}{\end{equation}}

\definecolor{mygreen}{rgb}{0,0.5,0}
\definecolor{myblue}{rgb}{0,0,0.75}
\definecolor{mymagenta}{cmyk}{0,1,0,0.12}

\begin{document}

\title{Disentangling Sources of Quantum Entanglement in Quench Dynamics}

\author{Lorenzo Pastori}
\affiliation{Institute of Theoretical Physics, Technische Universit\"at Dresden, 01062 Dresden, Germany}	
\author{Markus Heyl}
\affiliation{Max-Planck-Institut f\"ur Physik komplexer Systeme, N\"othnitzer Stra\ss e 38, 01187 Dresden, Germany}
\author{Jan Carl Budich}
\affiliation{Institute of Theoretical Physics, Technische Universit\"at Dresden, 01062 Dresden, Germany}

\date{\today}

\begin{abstract}
	Quantum entanglement may have various origins ranging from solely interaction-driven quantum correlations to single-particle effects.
	Here, we explore the dependence of entanglement on time-dependent single-particle basis transformations in fermionic quantum many-body systems, thus aiming at isolating single-particle sources of entanglement growth in quench dynamics.   
	Using exact diagonalization methods, for paradigmatic non-integrable models we compare to the standard real space cut various physically motivated bipartitions. 
	Moreover, we search for a minimal entanglement basis using local optimization algorithms, which at short to intermediate post-quench times yields a significant reduction of entanglement beyond a dynamical Hartree-Fock solution.
	In the long-time limit, we identify an asymptotic universality of entanglement for weakly interacting systems, as well as a cross-over from dominant real-space to momentum-space entanglement in Hubbard-models undergoing an interaction quench.
	Finally, we discuss the relevance of our findings for the development of tensor network based algorithms for quantum dynamics.
\end{abstract}

\date{\today}

\maketitle

Entanglement as one of the most fundamental traits of quantum mechanics plays a key role in various physical contexts, ranging from the 
characterization of complex many-body states \cite{WenBook,AmicoVedral_EntangRev} to its use as a resource for quantum technologies \cite{NielsenChuangBook,BennetEtAl_QuantumTeleportation,PlenioVedral_EntQInfo_Rev,Horodecki_x_4_QuantumEntanglement,MicroscopyEnhancement}.
On the other hand, strong entanglement in correlated quantum matter poses a significant challenge that limits both its analytical and numerical description.
This becomes particularly relevant in nonequilibrium dynamics, where entanglement generically grows rapidly with time \cite{LiebRobinson, CalabreseCardy_EntEvo1D, DeChiaraEtAl_EntEvoHeisenberg, LauchliKollath_EntQuench1Dbosons, KimHuse_BallisticEntang, Daley_EntGrowthQuench, Cotler_EntQuenchFieldTheory, Abanin_EntDynamics, LuitzBarLev, Mezei_EntSpreadChaotic, NahumEtAl_EntGrowthRandomUnitary, Calzona2017, Mistakidis2018, PollmannSondhi_PRXall, NahumEtAl_EntDyn1Drand, SuraceTagliacozzo}.
This growth, however, may originate from mechanisms of different inherent complexity:
While it can already be found in analytically solvable free theories, at the opposite end entanglement can be purely interaction-induced.
Yet, it has remained largely unexplored as to what extent a heterogeneous dynamical entanglement content may be distinguished according to its various sources on general grounds.

\begin{figure}[htp!]
	\includegraphics[width=\linewidth]{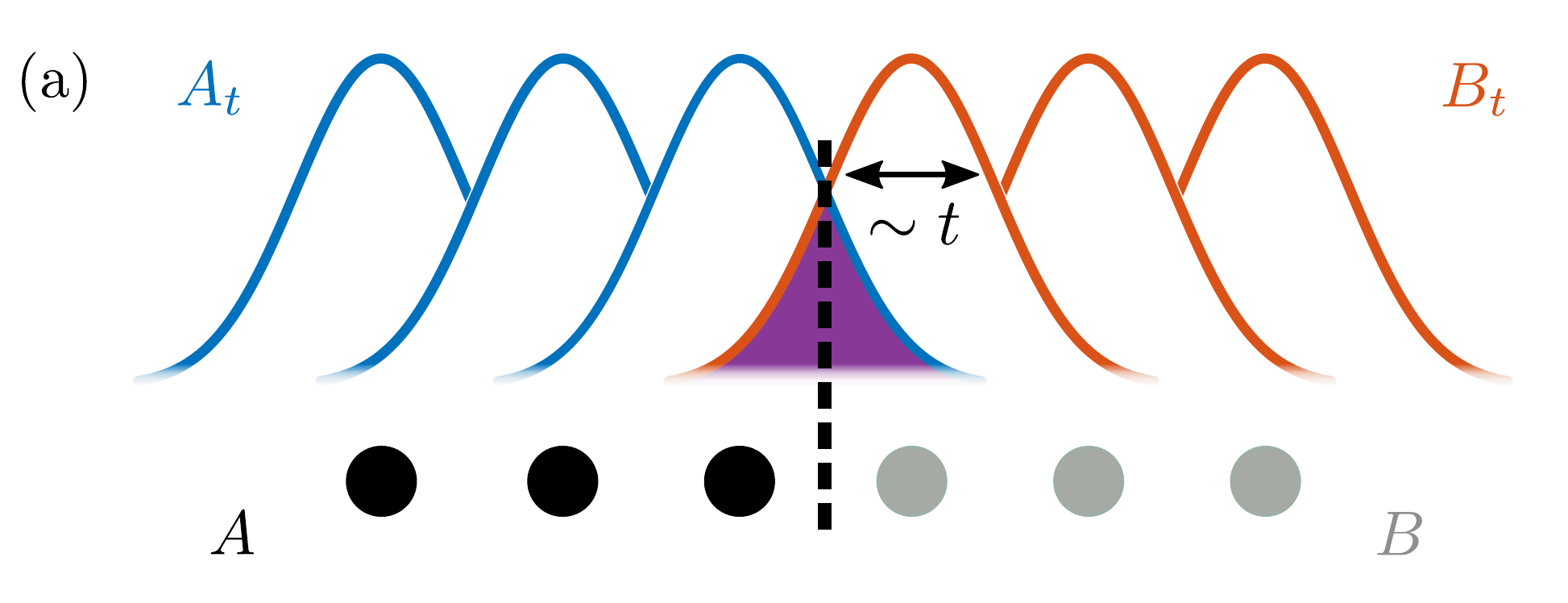}
	\includegraphics[width=\linewidth]{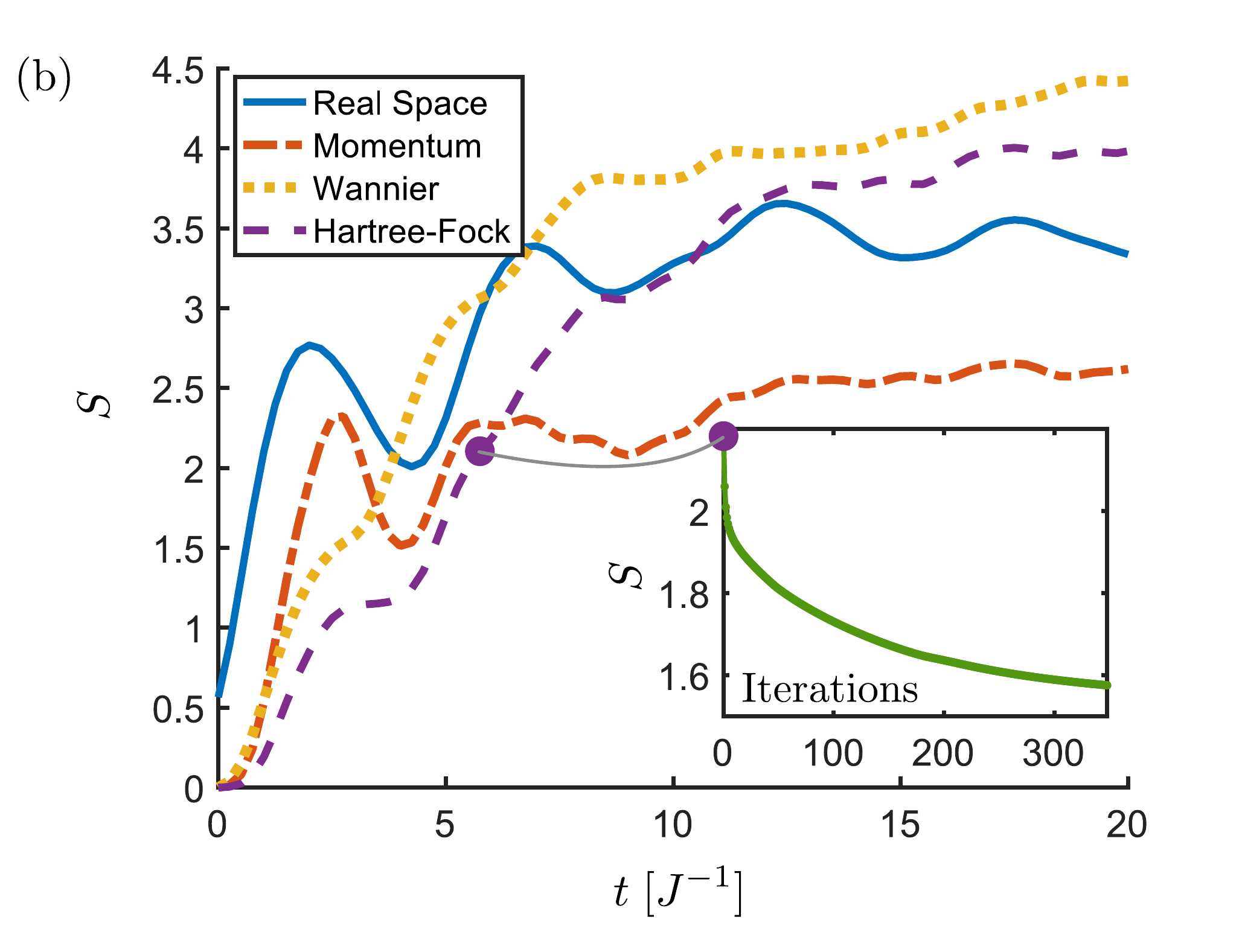}
	\caption{(a) Illustration of a time-dependent entanglement cut $A_t/B_t$ corresponding to a basis of dispersing Wannier orbitals, in comparison to a standard real-space cut $A/B$. (b) Time dependence of the entanglement entropy $S$ of the interacting model defined in Eqs.~(\ref{eqn:ssh}-\ref{eqn:HI}) far from equilibrium. Different curves correspond to different entanglement cuts (see plot legend). Inset: Basis-optimization of entanglement beyond the dynamical Hartree-Fock cut. Parameters are $L=24, n_p=12, U=J=1.0, W=0,m_i=1.8,m_f=0.2$.}
	\label{fig1}
\end{figure}

The purpose of this work is to investigate how the single-particle content of entanglement in quantum quench dynamics can be isolated from genuine many-body complexity, thereby revealing physically distinct sources of quantum correlations.
To this end, we study the dependence on generally {\textit{time-dependent}} single-particle basis rotations (entanglement-cuts) of the half-system entanglement entropy
\begin{equation}
S(t) = -\mathrm{tr} \left[ \rho_{A_t} \log\left( \rho_{A_t} \right) \right],
\label{eqn:se}
\end{equation}
where $\rho_{A_t} = \mathrm{tr}_{B_t} \rho$ denotes the reduced density matrix of subsystem $A_t$ at time $t$ obtained from the full density matrix $\rho=\lvert\psi(t) \rangle \langle \psi(t) \rvert$ by tracing out its complement $B_t$ (see Fig.~1(a)), and $\lvert\psi(t) \rangle$ is the state of the system at time $t$.
Going beyond the conventionally used real or momentum space basis, we investigate different choices of single-particle orbitals for the bipartition of the system in $A_t$ and $B_t$, including physically motivated states such as dispersing Wannier orbitals or time-dependent Hartree-Fock solutions. 
Furthermore, we search for an optimal (i.e. yielding minimal entanglement $S$) orbital basis using local optimization algorithms (see inset in Fig.~\ref{fig1}(b)), in order to separate genuine many-body entanglement content, originating from quantum correlations involving several particles, from single-particle contributions.

As we exemplify for several non-integrable fermionic quantum many-body systems, for quite long transient times the entanglement entropy indeed exhibits a significant basis-dependence (see Fig.~\ref{fig1}(b)).
Based on our concrete case studies in the framework of exact diagonalization, in this context we identify several general principles for both transient and asymptotically long times, including $(i)$ an asymptotic universality (in the sense of basis-independence) of entanglement in band insulators with weak to modest interactions, and $(ii)$ a cross-over between real-space and momentum-space entanglement in Hubbard models undergoing an interaction quench. 
Our results contribute towards the physical understanding of entanglement in complex quantum matter, and might be of key relevance for the development of numerical algorithms for quantum dynamics based on tensor network approaches \cite{WhiteDMRG,McCulloch_iDMRG,StellanMPS,SchollwoeckMPS,Daley2004,VerstraeteCiracMPS,HieidaVBS,VerstraeteCitracVBS,VerstraeteCiracPEPS,VerstraeteMurgCirac_revPEPS,VidalMERA1,VidalMERA2,ShiDuanVidalTTN,TagliacozzoEvenblyVidalTTN}, whose efficiency crucially depends on entanglement scaling.

{\emph{Entanglement dynamics from Krylov time propagation.}}
For our quantitative study, we consider a system of $n_p$ interacting fermions on a lattice with sites $j=1,\ldots, L$, which are annihilated by the vector of field operators $c=(c_1,\ldots,c_L)$. 
To quench the system out of equilibrium, the Hamiltonian is suddenly changed from $H_i$ to $H_f$ at time $t=0$, assuming that the many-body state $\lvert \psi(t=0)\rangle$ was prepared as the ground state of $H_i$ at $t<0$.
Using numerically exact Krylov time-propagation methods \cite{KrylovTimeProp, som}, for several examples of $H_i$ and $H_f$ representing quenches into non-integrable systems, we then compute the time-evolved wave function ($\hbar=1$)
\begin{align}
\lvert \psi(t)\rangle = \text{e}^{-i H_ft}\lvert\psi(0)\rangle.
\end{align}
In order to study the dynamics of the (von Neumann) entanglement entropy $S(t)$, we decompose the system into two subsystems, $A_t$ and $B_t$, each containing $L/2$ orbitals and on average $n_p/2$ particles. 
However, we do not restrict this decomposition (entanglement cut) to the real space lattice space, but consider time-dependent basis orbitals $\tilde c_t = u_t c$, resulting from the real space lattice representation by an arbitrary $\text{U}(L)$-transformation $u_t$ (with $\text{U}(L)$ denoting the group of $L\times L$ unitary matrices).
To represent the time evolved wave function in an arbitrary orbital basis, we need to express $u_t$ in many-body Hilbert space, where it is denoted by $U_t$. 
Naively applying $U_t$ to the real space representation  $\psi (t)$ of the state vector $\lvert \psi(t)\rangle$ would generate an unfeasible computational cost $\sim N^2$, where $N$ is the dimension of the many-body Hilbert space of $n_p$ particles.
However, exploiting that $U_t$ formally may be seen as a ``time-evolution" with respect to the the free Hamiltonian $\tilde H_t=\sum_{j,\ell} \left(i\log u_t\right)_{j,\ell}c_j^\dag c_{\ell} $ yields
\begin{align}
\tilde \psi (t)= U_t \psi(t) =\text{e}^{-i\tilde H_t}\psi(t),
\label{eqn:basistrafo}
\end{align} 
which naturally allows for the efficient ($O(N)$) computation of the transformed representation $\tilde \psi(t)$, again using Krylov time-progation with respect to the fictitious Hamiltonian $\tilde H$.

In the following, we present exact numerical data on the quench dynamics of systems with up to $L=24$ sites and $n_p=12$ fermions, where $N \approx 2.7\cdot10^6$.
Concretely, we study two examples of non-integrable ergodic model systems, $(i)$ a 1D two-band band insulator that is quenched to a topological insulator (TI) phase with weak to modest interactions, and $(ii)$ a metallic single-band Fermi-Hubbard model with next-nearest neighbor (NNN) interactions. 
By comparison of these physically quite different scenarios, we will identify various features relating to the basis-dependence of entanglement in quantum quench dynamics.

{\emph{$(i)$ Weakly interacting topological insulator.}}
We first discuss a one-dimensional (1D) two-banded system similar to the celebrated Su, Schrieffer, and Heeger (SSH) model \cite{SSH,SSHReview}.
There, the lattice index $j=1,\ldots L$ is decomposed into a unit-cell index $i=1,\ldots, L/2$ and a sublattice index $\alpha=a,b$ such that $j=2i-1$ for site $a$ in cell $i$ and $j=2i$ for site $b$ in cell $i$, respectively.
In reciprocal space, the two-band Bloch Hamiltonian at lattice momentum $k=4\pi p/L,~p=0,\ldots,L/2-1$ of the system is defined as
\begin{align}
h(k)= \left[m(t)+J \cos(k)\right]\sigma_x+ J \sin(k)\sigma_y,
\label{eqn:ssh}
\end{align}
where $\sigma_i$ are the standard Pauli matrices acting on $a/b$ sublattice space, $m(t)$ plays the role of a Dirac mass parameter that is quenched from $m_i$ to $m_f$ at $t=0$, and $J$ is the hopping strength between neighboring unit cells. 
For $\lvert m\rvert<\lvert J\rvert$, the model is in a topological insulator phase protected by the chiral symmetry $\sigma_zh(k)\sigma_z=-h(k)$, while at $\lvert m\rvert=\lvert J\rvert$ a topological quantum phase transition to a trivial band insulator phase extending over the parameter regime $\lvert m\rvert>\lvert J\rvert$  occurs. 
Here, we focus on quenches which change the topological phase of the system Hamiltonian, by choosing $m_i$ in the trivial, and $m_f$ in the non-trivial phase, respectively.

For the (standard) real space entanglement cut, even the non-interacting system is found to exhibit linear entanglement growth at small times.
However, this entanglement is readily seen to be entirely basis-dependent: The exact solution $\lvert \psi(t)\rangle$ stays a single Slater determinant at all times, leading to zero entanglement in a suitable basis obtained from time-evolved Wannier functions or Bloch functions representing the initial uncorrelated state.
This provides a conceptually simple example for how strongly the entanglement of a system far from equilibrium may depend on the chosen representation.

\begin{figure}[t]
	\includegraphics[width=\linewidth]{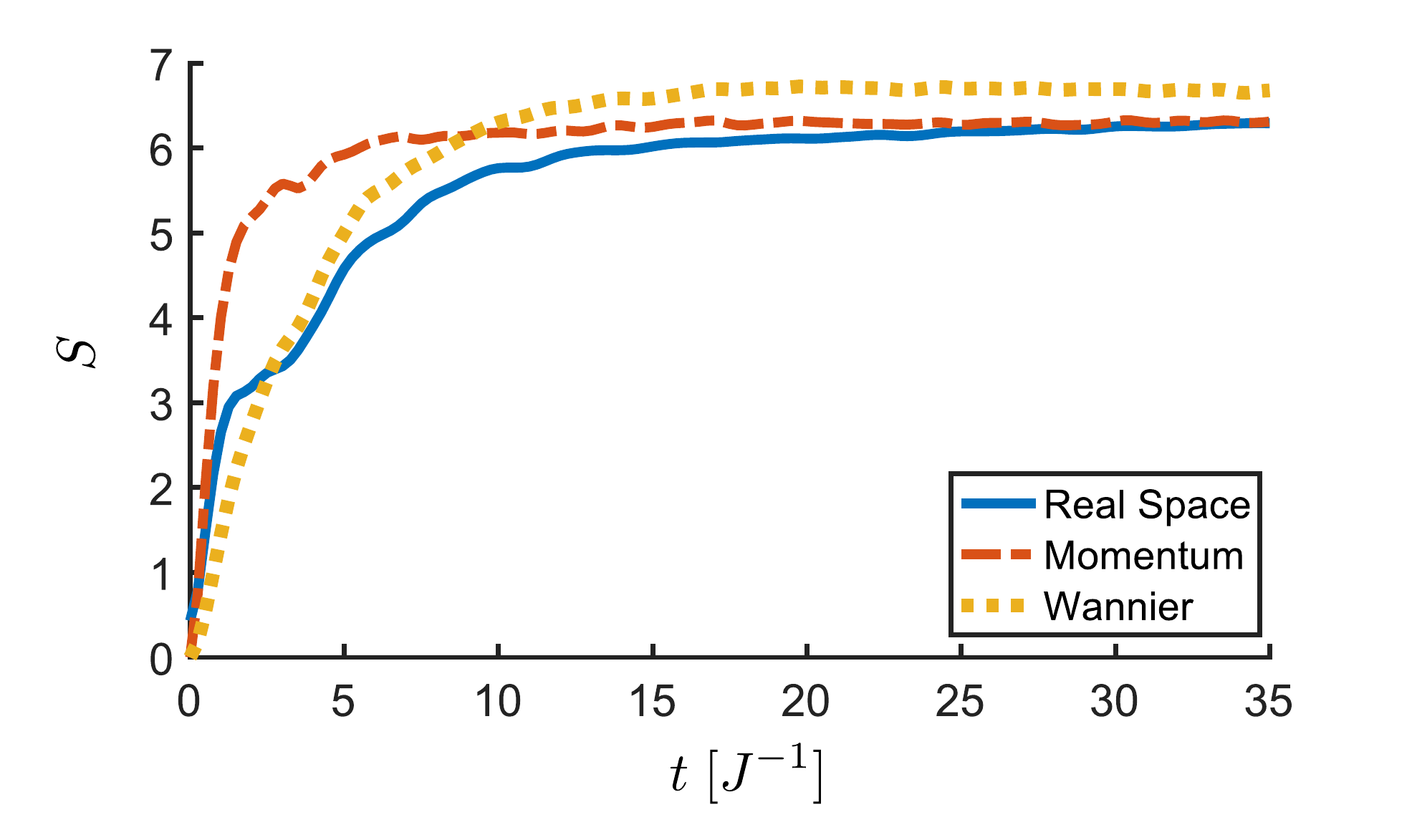}
	\caption{Asymptotic universality of the entanglement entropy $S$ of the equilibrating quenched interacting topological insulator model defined in Eqs.~(\ref{eqn:ssh}-\ref{eqn:HI}): Different curves corresponding to different entanglement cuts (see plot legend) saturate to similar values at long times. Parameters: $L=24, n_p=12, W=J=1.0, U=0,m_i=1.8,m_f=0.2$.}
	\label{fig2}
\end{figure}

We now address the natural question as to how an integrability breaking interaction term affects this phenomenology.
To this end, we add to the free model (\ref{eqn:ssh}) the interaction Hamiltonian
\begin{align}
H_I=\sum_i\Big[U n_{i,a}n_{i,b}+W(n_{i,a}n_{i+1,b}+n_{i,b}n_{i+1,a})\Big],
\label{eqn:HI}
\end{align}
with $n_{i,\alpha}=c^{\dagger}_{i,\alpha}c_{i,\alpha}$ ($\alpha=a,b$), where already a non-zero $U$ or $W$ individually is sufficient to make the model non-integrable, and both terms preserve the protecting chiral symmetry.
We first discuss the case on $U=J,W=0$ for transient times (see Fig.~\ref{fig1}(b)).
For the Bloch and Wannier entanglement cut, in which the non-interacting solution would exhibit zero entanglement, $S$ is lower than in the real space cut, for short to intermediate times.
Interestingly, evolving the Wannier basis with respect to a dynamical Hartree-Fock Hamiltonian, thus accounting for interactions at mean field level leads to a substantial reduction of $S$ for quite long times.
Searching for an optimal entanglement cut by treating the translation-invariant Hermitian matrix $\tilde H_t$ as a variational ansatz generating the basis transformation $U_t$ (see Eq.~(\ref{eqn:basistrafo})) \cite{som}, we are able to further reduce $S$ significantly below the dynamical Hartree-Fock basis (see inset of Fig.~\ref{fig1}(b)) by using local optimization algorithms such as Gradient Descent and ADAM \cite{AdamPaper}. 
We now switch on $W>0$ which is generally found to speed up the process of thermalization in 1D topological insulator models \cite{AndiPRB}, thus allowing us to look at the long-time (close to equilibration) behavior of our model.
For weak to modest interactions, we observe a quite remarkable universal aspect of entanglement in the long-time limit, namely that $S$ becomes largely independent of the choice of basis (see Fig.~\ref{fig2}).
This behavior exemplifies a quite generic mechanism that can be understood with the following physical picture \cite{som}. 
When a system is quenched far from equilibrium by changing its gapped single particle Hamiltonian, weak interactions are expected to lead to thermalization at an effective temperature $\beta_e$ of the system with respect to the close to non-interacting post-quench Hamiltonian.
However the thermal entropy of a free Hamiltonian (such as Eq.~(\ref{eqn:ssh})) generically is independent (up to non-extensive boundary effects) on the basis of a bipartition with an average of $n_p/2$ particles in each subsystem.
Thus, the basis-dependence of the entanglement entropy in this scenario is a transient effect, even though it might last to long times.
This basis-independence in weakly-interacting thermalizing insulators also affects the margin to be gained by entanglement minimization: By optimizing the aforementioned basis-change Hamiltonian $\tilde H_t$, starting from the set of Wannier orbitals in the long-time limit, here it is not possible to significantly reduce the entanglement entropy.

\begin{figure}[t]
	\includegraphics[width=\linewidth]{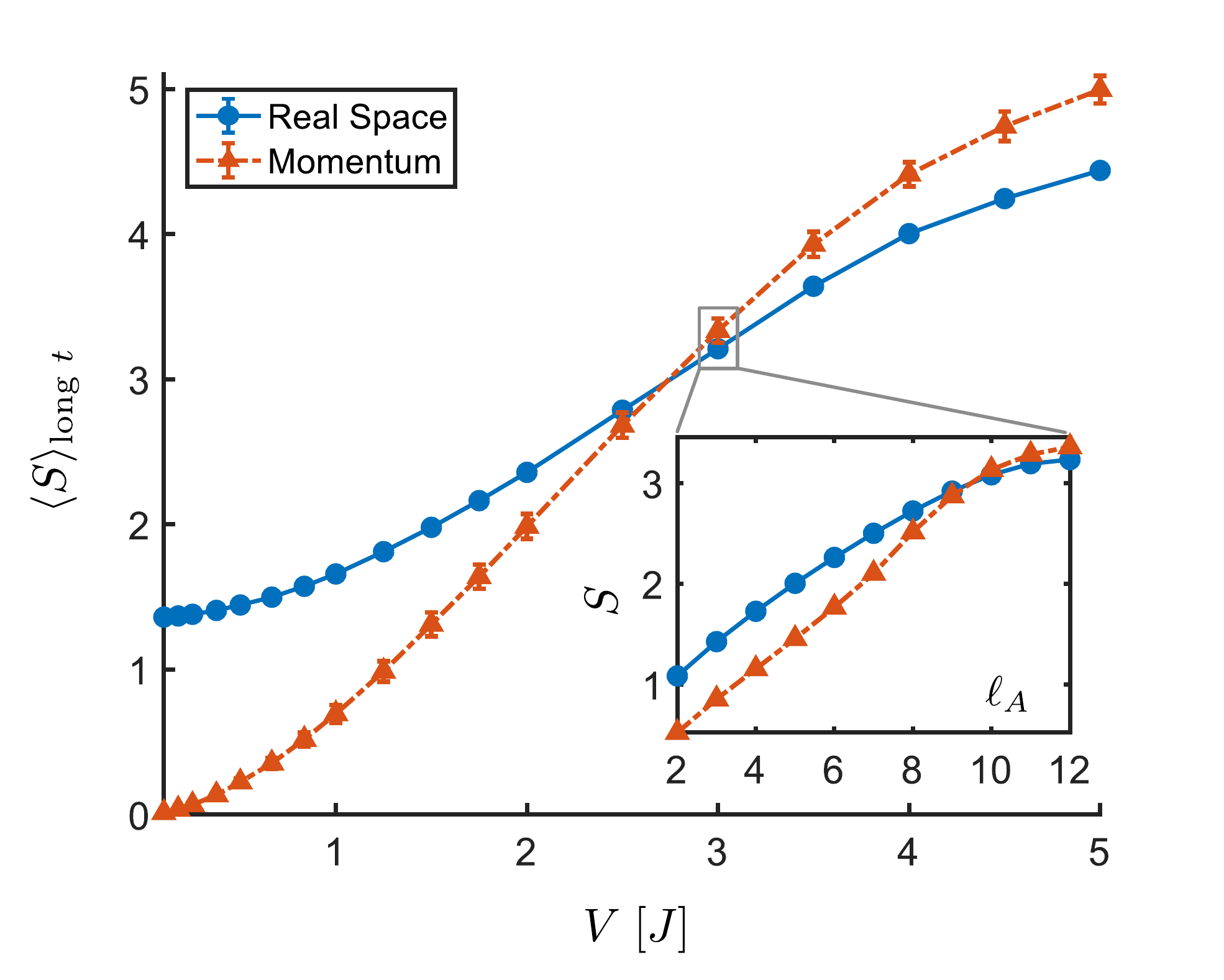}
	\caption{Long-time average of $S$ for the spinless Hubbard model (\ref{eqn:NNNHubbard}) with $L=24,~\nu=1/3$ as a function of the post-quench next nearest neighbor interaction strength $V$. Inset: Scaling of $S$ with the number of orbitals $\ell_A$ in $A$.}
	\label{fig3}
\end{figure}

{\emph{$(ii)$ Spinless NNN Fermi-Hubbard model.}} 
As a second non-integrable model system, we consider a spinless single band Fermi-Hubbard model with NNN interaction at filling $\nu=n_p/L$, defined by the Hamiltonian
\begin{align}
H=\sum_{j}\left[-J(c_j^\dag c_{j+1} + \text{h.c.})+V(t)(n_j-\nu)(n_{j+2}-\nu)\right],
\label{eqn:NNNHubbard}
\end{align}
where $J$ (fixed to $1$ in the simulations) denotes the hopping strength, $n_j=c_j^\dag c_j$, and $V(t)\ge 0$ represents the repulsive NNN interaction strength which is suddenly changed from $0$ to a finite value over the quench at $t=0$.
We focus on $\nu =1/3$ in the following, where the model (\ref{eqn:NNNHubbard}) at zero temperature is known to stay in a metallic Luttinger liquid phase up to large $V>0$ \cite{NNNphasediag_PRB}, and has been found to exhibit strongly ergodic behavior \cite{NNNergodic_PRE}. 
When comparing the long-time entanglement entropy of this model in real space and momentum space, we find an interesting basis-dependence that persists even in the long-time limit: For weak to modest $V$, the long-time entanglement in momentum space is lower than in real space, whereas at strong correlations ($V\gtrsim 3 J$), the real space cut leads to lower entanglement entropy (see Fig.~\ref{fig3}).
This behavior may be interpreted as a dynamical manifestation of the generic competition between an interaction term that is diagonal in real space and a free band structure that is diagonal in momentum space in Hubbard models \cite{som}. 
At short times, by contrast, entanglement for all $V$ grows steeper in the momentum cut, owing to the non-local nature of the interaction in reciprocal space.
In both cuts, entanglement clearly exhibits extensive (volume law) scaling for long times, as expected for an equilibrating ergodic system (see inset of Fig.~\ref{fig3}).

{\emph{Concluding Discussion.}
	We conclude by putting our present results into a broader perspective from several angles.
	Generally speaking, non-integrable ergodic systems are known to exhibit a generic dynamical behavior for the entanglement entropy $S$ in coherent quench dynamics starting from a weakly entangled (equilibrium) state: 
	$S$ initially grows linearly in time reflecting a ballistic spreading of quantum information at a certain velocity \cite{LiebRobinson, CalabreseCardy_EntEvo1D, DeChiaraEtAl_EntEvoHeisenberg, LauchliKollath_EntQuench1Dbosons, KimHuse_BallisticEntang, Daley_EntGrowthQuench, Cotler_EntQuenchFieldTheory, Abanin_EntDynamics, LuitzBarLev, Mezei_EntSpreadChaotic, NahumEtAl_EntGrowthRandomUnitary, Calzona2017, Mistakidis2018, PollmannSondhi_PRXall, NahumEtAl_EntDyn1Drand, SuraceTagliacozzo}.
	At large times, close to thermalization, $S$ approaches an extensive value (volume law), i.e. $S=\alpha V_A$, where $V_A$ denotes the volume of subsystem $A$, since the entropy as an equilibrium thermodynamic potential is always extensive.
	While all of our present findings fit into this rough generic framework, here we identified and microscopically exemplified several general principles relating to the basis dependence and heterogeneous physical nature of dynamical entanglement growth.

	First, for insulators with weak to modestly strong interactions that are quenched out of equilibrium by a change in their band structure, the transient entanglement exhibits very strong basis dependence and may be reduced significantly beyond a dynamical mean field solution by means of single-particle basis rotations.
	By contrast, the volume-law coefficient $\alpha$ of the long-time average of $S$ becomes basis independent. 
	This is because the system is thermalizing with respect to a nearly free Hamiltonian, and the thermal entropy for a free fermionic system typically is largely independent of the choice of bipartition, as long as each subsystem contains half of the degrees of freedom and accommodates on average half of the particles.
	
	Second, in Hubbard models that are undergoing an interaction quench leaving the non-interacting Hamiltonian unchanged, a significant basis-dependence of entanglement persists in the long-time limit, where the considered ergodic systems exhibit basis-dependent extensive entanglement scaling:
	Up to intermediate post-quench interaction strengths entanglement is found to be weaker in reciprocal space while at stronger correlations the real-space cut has lower in $S$.
	This behavior dynamically reflects the competition between kinetic energy and interactions characteristic for Hubbard models.

	These findings on distinguishing single-particle contributions to $S$ from inherently more complex quantum correlations are not only of fundamental interest, but might also serve as important guidelines for the pursuit of devising more efficient numerical algorithms for quantum dynamics.
	Within the realm of thermal equilibrium, the observation that low-temperature states typically exhibit significantly lower (area-law \cite{Hastings_AreaLaw1D,Eisert_AreaLawReview}) entanglement than generic states has been crucial for systematically taming the exponential complexity of correlated systems, e.g. with the advent of tensor network methods \cite{WhiteDMRG,McCulloch_iDMRG,StellanMPS,Daley2004,SchollwoeckMPS,VerstraeteCiracMPS,HieidaVBS,VerstraeteCitracVBS,VerstraeteCiracPEPS,VerstraeteMurgCirac_revPEPS,VidalMERA1,VidalMERA2,ShiDuanVidalTTN,TagliacozzoEvenblyVidalTTN}, where the basis-dependence of entanglement has been successfully exploited in the context of spin-boson models \cite{Weichselbaum2012} and quantum chemistry settings \cite{Mitrushenkov2001, Szalay2015, Krumnow2016}. 
	However, far from equilibrium, generically the dynamical proliferation of entanglement eventually builds up an exponential wall even for the most powerful known computational methods. 
	In this context, our results provide systematic insights addressing the question as to what extent both transient entanglement growth and long-time entanglement may be made computationally manageable, e.g. by extending tensor network methods to a flexible representation of the physical degrees of freedom that allows for time-dependent basis choices.  
	
	In this work, we have been concerned with the influence of {\emph{single-particle}} basis rotations on entanglement dynamics.
	At a conceptual level, our search for an optimal basis in which $S$ is lowest may be interpreted as entanglement-based dynamical mean-field approach aimed at minimizing the single-particle contributions to $S$.
	Thinking further along these lines, it is readily conceivable to consider more elaborate approximate solutions based on genuine many-body methods as a starting point for the choice of a {\textit{correlated}} basis, which may eliminate even some higher-order quantum correlations from the system.
	Quantitatively exploring this approach and constructing hybrid numerical methods, where a physically motivated approximate solution serves as the starting point to be augmented by an unbiased tensor-network simulation, represents an interesting subject of future work.
	
	\emph{Note added. } While preparing this manuscript for submission, two related preprints appeared online \cite{eisert_basisMPS,transport_freqdomain}. The authors of Ref.~\cite{eisert_basisMPS} have applied the idea of a dynamical single-particle basis rotations on MPS, optimizing at each step a sequence of unitary two-site gates acting on the MPS. In Ref.~\cite{transport_freqdomain}, the authors have chosen the frequency basis as a computationally advantageous single-particle basis for a MPS transport calculation.
	
	\emph{Acknowledgements. } We acknowledge discussions with S. Barbarino, M. Hermanns, C. Mendl, and H.-H. Tu. L.P. and J.C.B. acknowledge financial support from the DFG through SFB 1143 (Project No. 247310070). Our numerical calculations were performed on resources at the TU Dresden Center for Information Services and High Performance Computing (ZIH).

	\bibliographystyle{apsrev}


\ \\ \ \\ \ \\ \ \\

\begin{center}
	{\bf APPENDIX}
\end{center}

\setcounter{equation}{0}
\setcounter{figure}{0}

\appendix

\def\bra#1{\langle #1 |}  
\def\ket#1{| #1 \rangle}  
\def\d {^{\dagger}}  
\def\up {\uparrow}  
\def\dn {\downarrow}  

\renewcommand{\theequation}{A\arabic{equation}}
\renewcommand{\thefigure}{A\arabic{figure}}
\renewcommand{\bibnumfmt}[1]{[A#1]}
\renewcommand{\citenumfont}[1]{A#1}

\section{Krylov Time Propagation}
Here we briefly elaborate on the Krylov time propagation method \cite{SOMNautsWyatt_KrylovTimeProp}, used in the main text for the time-evolution as well as for the single-particle basis rotation of the many-body state $\lvert \psi\rangle$. We start with the case of conventional time-evolution, where the goal is to calculate $\ket{\psi(\Delta t)} = \text{e}^{-iH\Delta t}\ket{\psi}$ from a known initial state $\ket{\psi}$, with a Hamiltonian $H$ that is assumed to be time-independent. The idea of Krylov time propagation is based on the observation that the time-evolved state $\ket{\psi(\Delta t)}$, up to a given order $N_{\text{K}}-1$ in $\Delta t$ reads as
\begin{equation}
\ket{\psi(\Delta t)}\approx\sum_{m=0}^{N_{\text{K}}-1}\frac{(-i\,\Delta t)^m}{m!}\,H^m\ket{\psi}\,\,,
\end{equation}
and thus manifestly belongs to the $N_{\text{K}}$-th Krylov subspace $\mathcal{K}_{N_{\text{K}}}(H;\ket{\psi}) = \text{span}\big\{\ket{\psi},H\ket{\psi},H^2\ket{\psi},...\,,H^{N_{\text{K}}-1}\ket{\psi}\big\}$. Using this insight, the Hamiltonian $H$ may be expressed as a tri-diagonal matrix $T$ in the $N_{\text{K}}$-dimensional Krylov subspace, where typically $N_{\text{K}}\approx 50$. Then, the matrix exponential $\text{e}^{-iT\Delta t}$ may readily be expressed and explicitly applied to $\ket{\psi}$ within this Krylov subspace, before transforming the resulting vector back to the full Hilbert space. 

Practically, one needs to generate a basis for $\mathcal{K}_{N_{\text{K}}}(H;\ket{\psi})$, which can be done using the Lanczos method (see e.g.\ Ref.~\cite{SOMLinGubernatis_ED}). We denote with $Q$ the matrix whose columns are the $N_{\text{K}}$ orthonormal basis states $\ket{q_n}$ (with $n=0,...,N_{\text{K}}-1$) for the Krylov subspace, where $\ket{q_0}=\ket{\psi}$. The initial state $\ket{\psi}$ is thus represented in the Krylov subspace by the $N_{\text{K}}$-component vector $Q\d\ket{\psi}=(1 \; 0 \; ... \; 0)^{\top}$. The vector $\text{e}^{-iT\Delta t}Q\d\ket{\psi}$ therefore represents the time-evolved state $\ket{\psi(\Delta t)}$ in $\mathcal{K}_{N_{\text{K}}}(H;\ket{\psi})$ truncated to $O(\Delta t^{N_{\text{K}}-1})$. Hence, the full time-evolved state is retrieved from
\begin{equation}
\ket{\psi(\Delta t)}=Q\,\text{e}^{-iT\Delta t}\,Q\d\ket{\psi}+O(\Delta t^{N_{\text{K}}}) \,\,,
\end{equation}
at an overall computational cost of $O(N_{\text{K}}\, N)$, where $N$ denotes the dimension of the many-body Hilbert space.

To apply this method also to basis transformations, we note that it can efficiently perform any unitary transformation $U=\text{e}^{-i\tilde{H}}$ on the state $\ket{\psi}$, provided that the Hermitian many-body operator $\tilde{H}$ can be efficiently applied to a state vector. Turning to the single particle transformations $\tilde c = u_t c$ with $u_t\in\text{U}(L)$ used in the main text, the Hermitian many-body operator $\tilde{H_t}$ can be constructed as the following fictitious non-interacting Hamiltonian $\tilde{H_t}=\sum_{j,\ell}(i\log u_t)_{j,\ell}c_j\d c_{\ell}$, where the $c_j$ denote the annihilation operators for the $j$-th orbital. This reduces the intuitive cost of $O(N^2)$ for a unitary basis transformation to the lower computational cost of $O(N)$ that is typical of Krylov methods, thus removing the computational bottleneck of our simulations from the application of basis transformations.

\section{Entanglement Optimization}
Here we provide some details on the variational optimization of the entanglement entropy $S$ used in the main text. The goal is to find an optimal single-particle basis transformation $u\in\text{U}(L)$ yielding the minimum entanglement entropy for a equal-size bipartition of the resulting single-particle modes $(\tilde c_1,\ldots \tilde c_L)$, where we dropped the time-dependence of all involved quantities for notational brevity. For a given set of $L$ single-particle modes corresponding to $u$, we denote with $\tilde{A}$ and $\tilde{B}$ the two subsystems containing $L/2$ modes each. The (von Neumann) entanglement entropy for $\tilde{A}$ is then written as
\begin{equation}
S(\rho_{\tilde{A}}) = -\text{tr}[\rho_{\tilde{A}}\log\rho_{\tilde{A}}]\,\, ,
\end{equation}
where $\rho_{\tilde{A}}=\text{tr}_{\tilde{B}}\ket{{\psi}}\bra{{\psi}}$. As mentioned in the main text the representation (coordinate) vector $\psi = (\psi_1,\ldots, \psi_N)^{\top}$ of the state $\lvert \psi\rangle$ transforms under the basis rotation as
\begin{equation}
\tilde \psi = \text{e}^{-i\tilde{H}}\psi\,\,,
\end{equation}
where again $\tilde{H}=\sum_{j,\ell}(i\log u)_{j,\ell}c_j\d c_{\ell}$. For our variational minimization of $S(\rho_{\tilde{A}})$, we interpret the elements $(i\log u)_{j,\ell}\equiv\tilde{h}_{j,\ell}$ as fictitious hopping amplitudes of $\tilde{H}$, and use them as \emph{variational parameters}. We furthermore would like to impose the constraint that each bipartition should contain on average $n_p/2$ particles, i.e. half of the total number of particles $n_p$. To this end we minimize the \emph{entanglement cost function}
\begin{equation}
C(\rho_{\tilde{A}}) = S(\rho_{\tilde{A}})+\lambda\big|n_{\tilde{A}}-n_p/2\big|\,\,,
\label{eqn:ecost}
\end{equation}
where $n_{\tilde{A}}=\text{tr}(N_{\tilde{A}}\rho_{\tilde{A}})$ is the expectation value of the particle number operator $N_{\tilde{A}}$ in subsystem $\tilde{A}$. The second term with $\lambda>0$ in the above equation punishes the entanglement cuts resulting in an imbalanced particle number between subsystems $\tilde{A}$ and $\tilde{B}$, where we choose $\lambda\approx 40$ to basically obtain a hard constraint for practical purposes.

\begin{figure} [t]  
	\includegraphics[width=\linewidth]{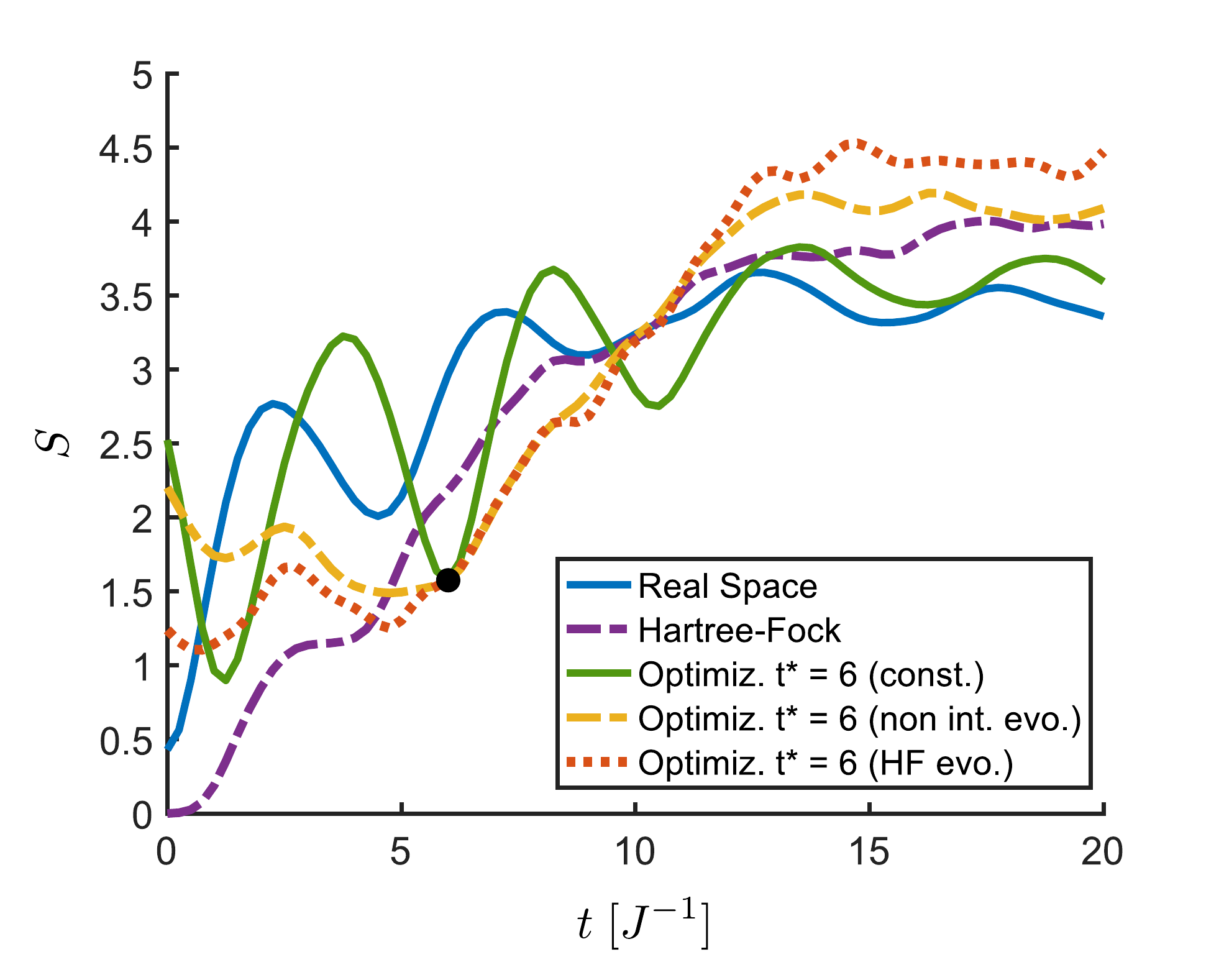}
	\caption{Time dependence of the entanglement entropy $S$ of the model defined in Eqs.~(4-5) in the main text, for $L=24, n_p=12, U=J=1.0, W=0,m_i=1.8,m_f=0.2$. The black dot denotes the value of entanglement for the bipartition of the optimized single-particle basis set at time $t^* = 6J^{-1}$ (see inset of Fig.~1(b) in the main text).}
	\label{figS1}
\end{figure}

For the minimization we adopt gradient based methods such as simple steepest descent or the Adam optimization algorithm \cite{SOMAdamPaper}. 
Such gradient methods require the calculation of the derivatives of $C(\rho_{\tilde{A}})$ with respect to all the parameters $\tilde{h}_{j,\ell}$, which we do numerically to first order.
Imposing translation invariance on $\tilde{h}_{j,\ell}$ automatically leads to Wannier-like (i.e. shift-orthogonal) single-particle orbitals, and reduces the number of real variational parameters from $O(L^2)$ to $O(L)$.
Denoting by $T$ the translation operator on the lattice, this ansatz limits our search to single-particle basis sets satisfying $T\,u\,T\d = u$.
We point out here that this choice excludes the physically relevant basis choice of Bloch states that are eigenstates of the translation operator $T$, satisfying $T\,u=u\,D$ with $D$ a diagonal matrix of phase factors.
However, our simulations using the unrestricted set of parameters $\tilde{h}_{j,\ell}$ for small systems indicate that the momentum cut of Bloch states always constitutes an isolated local minimum for the entanglement cost function (\ref{eqn:ecost}). 
Hence, for gradient-based optimization methods, the minimization within the space of Wannier-like orbitals, where the ansatz of translation invariance applies, is more promising and indeed found to be more successful. 

We point out that the translation operator $T$ in real space is different for the two models studied in the main text. In the case of the SSH model, $T$ translates by one unit-cell, which consists of two orbitals, hence the ansatz $T\,\tilde{h}\,T\d = \tilde{h}$ yields $2L$ independent variational parameters. In the case of the spinless Hubbard model with NNN interaction, $T$ translates by one physical site with a single orbital, hence resulting in an ansatz depending on $L$ independent variational parameters.

In Fig.~\ref{figS1} we show the time-evolution of the entanglement entropy for a bipartition of the optimized single-particle modes $u_*$ obtained from the dynamical Hartree-Fock solution at a fixed time $t^*$ of the post-quench time evolution (see inset of Fig.~1(b) in the main text). The black dot at $t^* = 6J^{-1}$ and $S\approx 1.57$ denotes the value of entanglement for this optimized set. The green solid curve represents the time-evolution of the entanglement obtained by keeping this single-particle basis set constant during the post-quench dynamics. The dashed yellow curve represents the entanglement between the optimized modes $u_*$ time-evolved with the post-quench single-particle Hamiltonian, whereas the dotted orange curve refers to their time-evolution with a dynamical Hartree-Fock Hamiltonian. We notice that $u_*$, and its time-evolved counterparts, yields a minimum for the entanglement entropy only at instant $t^*$ and in a small neighborhood of $t^*$. At further times the minimal-entanglement orbitals are generally different, highlighting the dynamical character of the optimal entanglement bipartition, i.e. the need to frequently update this orbital basis over time. The entanglement time evolution for constant $u_*$ features  oscillations similar to the ones of the entanglement in the (constant) real-space basis, whose period is consistent with the value of the post-quench single-particle band-gap. Time-evolving $u_*$ after time $t^*$ with either the bare non-interacting or the dynamical Hartree-Fock Hamiltonian seems to yield better results at short times than the case of constant $u_*$, in terms of entanglement minimization.

\begin{figure} [hb] 
	\includegraphics[width=\linewidth]{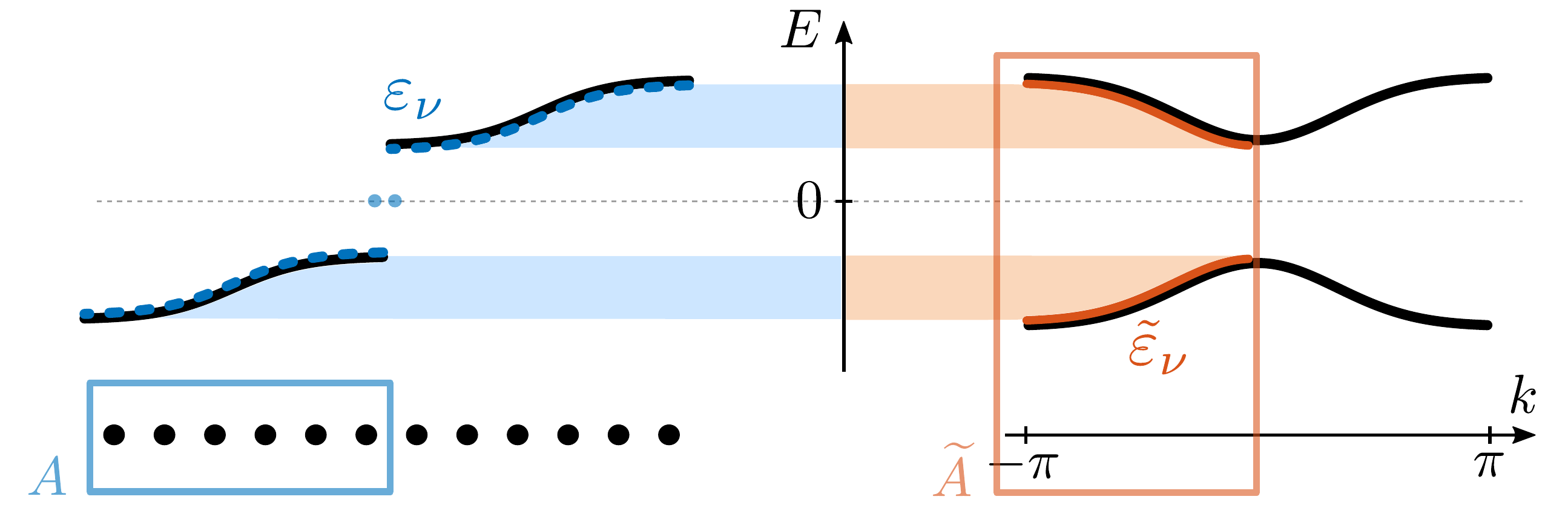}
	\caption{Schematic representation of the equivalence of the spectra of real-space and momentum restrictions of $H_f$. The single-particle spectrum of $H_f$ for the total system is shown by the thick black lines, on the right labeled by the momentum $k$. For the real-space cut $A$ (region in the blue box) the spectrum $\{\varepsilon_{\nu}\}$ of $H_f|_A$ is shown by the dashed blue line. For the momentum cut $\tilde{A}$, the spectrum $\{\tilde{\varepsilon}_{\nu}\}$ of $H_f|_{\tilde{A}}$ is shown by the orange thick lines. The two spectra concur, up to boundary modes (the two light-blue dots in the spectrum on the left), and corrections that vanish in the thermodynamic limit.}
	\label{figS2}
\end{figure}

\section{Asymptotic Basis-Independence of Entanglement Entropy}
In this section, we briefly elaborate on the long-time basis-independence of the entanglement entropy observed in the post-quench dynamics of the thermalizing SSH model. As explained in the main text, when the band structure of the system is suddenly changed by quenching its single-particle Hamiltonian from $H_i$ to $H_f$, the presence of weak interactions is generically expected to induce thermalization of the system, as intuitively the scattering processes will equilibrate the single-particle excitation created during the quench protocol \cite{SOMAndiPRB}. 

In the weakly interacting regime we can hence assume that in the long-time limit the system to good approximation thermalizes with respect to the \emph{free} part of the post-quench Hamiltonian $H_f$, in the sense that a subsystem $A$ would be described by a reduced density operator of the form
\begin{equation}
\rho_A = \frac{1}{Z_A}\,\text{e}^{-\beta_eH_f|_A} \,\,,
\end{equation}
where $\beta_e$ denotes the inverse temperature, generally depending on the density of excitations created during the quench, $H_f|_A$ denotes the restriction to subsystem $A$ of the free post-quench Hamiltonian $H_f$, and $Z_A=\text{tr}\big\{\text{e}^{-\beta_eH_f|_A}\big\}$. Since $H_f|_A$ is a free fermionic Hamiltonian, the (von Neumann) entanglement entropy $S(\rho_{A}) = -\text{tr}[\rho_{A}\log\rho_{A}]$ can be calculated as the thermal entropy of a free Fermi gas, which reads as (see e.g. Refs.~\cite{SOMPeschel_RedDensMatFree,SOMVidalEtAl_EntangCriticalSys})
\begin{equation}
S(\rho_{A}) = \sum_{\nu}\Big[\log\big(1+\text{e}^{-\beta_e\varepsilon_{\nu}}\big)+\,\frac{\beta_e\varepsilon_{\nu}}{1+\text{e}^{\beta_e\varepsilon_{\nu}}}\Big]
\label{eq:ThermalEntropyFermi}
\end{equation}
where the $\varepsilon_{\nu}$ denote the single-particle energies of $H_f|_A$. 

\begin{figure} 
	\includegraphics[width=\linewidth]{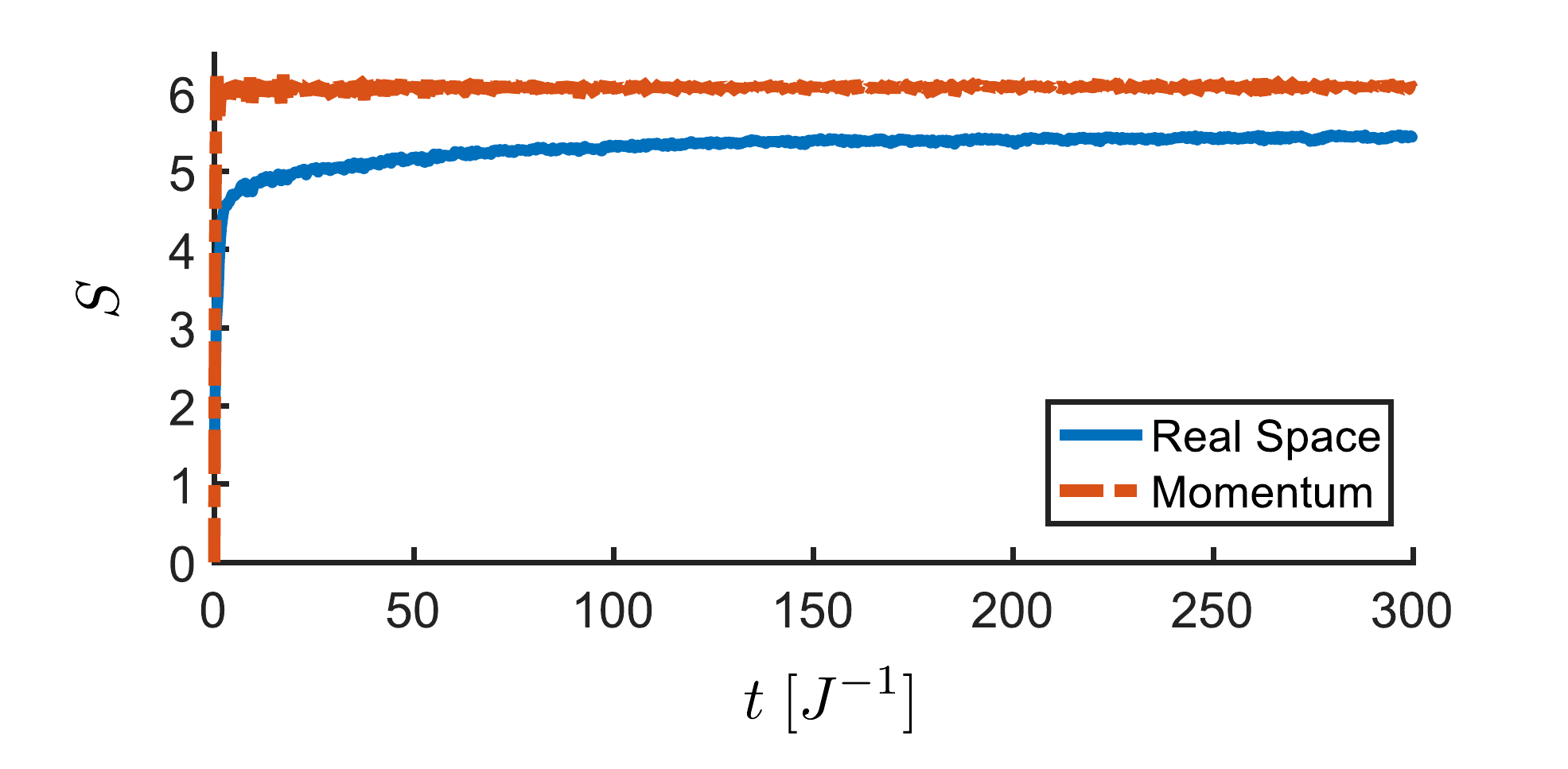}
	\caption{Time evolution of real-space (blue) and momentum (orange) entanglement entropy $S$ in the Fermi-Hubbard model with quenched NNN interaction defined in Eq.~(6) of the main text. Parameters used are $L=24, \nu=1/3, J=1$ and post-quench $V=20$.}
	\label{figS3}
\end{figure}

Since the above expression of $S(\rho_{A})$ depends only on the single-particle eigenvalues $\varepsilon_{\nu}$, we argue in the following that for a broad class of Hamiltonians $H_f$ the value of $S(\rho_{A})$ is independent of the choice of the subsystem $A$ up to non-extensive boundary terms, as long as $A$ has the same size and contains, on average, half of the particles. 
Specifically, let $H_f$ be the Hamiltonian of a system with an inversion symmetry $\mathcal{I}$ such that the Bloch Hamiltonian $h_f(k)$ in momentum space satisfies $\mathcal{I}\,h_f(k)\,\mathcal{I}\d = h_f(-k)$, hence its spectrum is symmetric with respect to $k=0$. 
As an example of quite different entanglement cuts, consider the subsystem $A$ as of consisting of half the system in real space, and $\tilde{A}$ a momentum space partition comprising only Bloch states in the momentum interval $I_{\tilde{A}}=[-\pi,0)$. 
Notice that this is the case for the SSH Hamiltonian and the entanglement cuts studied in the main text.
If the system is thermalized with respect to $H_f$, then the reduced density operator for $\tilde{A}$ reads as $\rho_{\tilde{A}} = Z_{\tilde{A}}^{-1}\,\text{e}^{-\beta_eH_f|_{\tilde{A}}}$. 
The crucial observation is that, under the above assumptions, the spectra of $H_f|_A$ and $H_f|_{\tilde{A}}$ are the same up to non-extensive boundary terms (e.g. topological edge states at the boundaries of $A$) and other corrections which vanish in the thermodynamic limit, provided that $A$ and $\tilde{A}$ contain the same number of single-particle modes and, on average, the same number of particles.
This is schematically shown in Fig.~\ref{figS2}.
Hence the entanglement entropies $S(\rho_{A})$ and  $S(\rho_{\tilde{A}})$ calculated from Eq.~(\ref{eq:ThermalEntropyFermi}) are equivalent, up to terms that do not scale with the size of the subsystems.
Similar statements hold when comparing to a cut of generic Wannier-like functions.
This explicates the asymptotic basis independence found in the main text for the quenched weakly interacting SSH model, and shows how it can be generalized to other weakly interacting quenched tight-binding models.  
This explanation is also corroborated by the fact that we did not observe any significant improvement, in terms of entanglement minimization with the translation-invariant ansatz introduced in the previous section, when trying to optimize the time-evolved basis sets in the long-time limit of the thermalizing post-quench dynamics.

\section{Real-Space vs. Momentum Entanglement Crossover in Spinless Hubbard Models}
In this section we briefly discuss the robustness of the crossover between dominant real-space entanglement to dominant momentum-cut entanglement, observed in the long time dynamics of the spinless Fermi-Hubbard model with next nearest neighbor (NNN) interactions studied in the main text (see Eq.~(6) and Fig.~3). We will show that the presence of such crossover is independent of the direction of the quench, and also robust to the addition of a nearest neighbor (NN) interaction term. These observations corroborate our intuitive explanation of this effect in terms of dynamical manifestation of the competition between hopping terms (diagonal in momentum representation) and interaction terms (diagonal in real-space representation).

We first observe that, in the interaction quench studied in the main paper, the momentum-cut entanglement remains dominant also for very strong values of the post-quench interaction. Figure \ref{figS3} shows a typical time-evolution of the entanglement for a quench in the strongly-interacting regime, where the entanglement for the momentum bipartition is significantly higher than the one for a real-space cut. We point out that in the case of real-space entanglement, the time at which the long-time limit is attained, i.e. when $S$ has reached its stationary value, increases with increasing interaction strength.

\begin{figure} 
	\includegraphics[width=\linewidth]{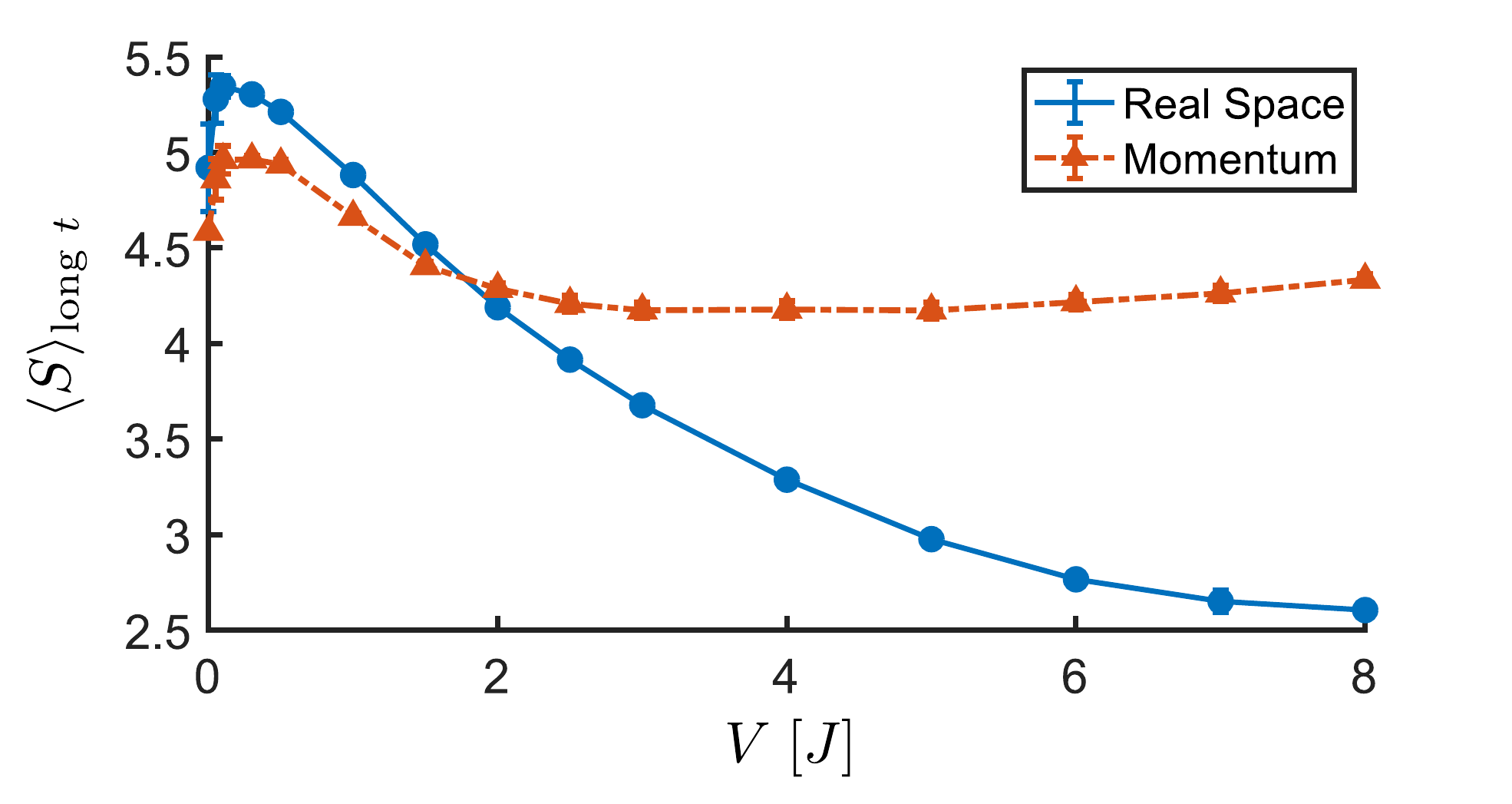}
	\caption{Long time average of $S$ for a reversed interaction quench in the spinless Hubbard defined in Eq.~(6) of the main text, with $L = 24$ and $\nu = 1/3$, as a function of the post-quench NNN interaction strength $V$.  The pre-quench value of the NNN interaction is $V_{\text{initial}}=10J$.}
	\label{figS4}
\end{figure}

We then investigated the long time behavior of the real-space and momentum-cut entanglement in the same model for the case of a quench in the opposite direction, namely from a strongly interacting regime (with $V=10J$) to weak-to-moderate interaction strengths. This is shown in Fig.~\ref{figS4}. We observe that also in this case the long time behavior shows a crossover from dominant real-space to dominant momentum-cut entanglement, for increasing value of the post-quench NNN interaction.

We also investigated the robustness of this crossover to the presence of a NN interaction term, studying interaction quenches in the following model:
\begin{align}
H=&\sum_{j}\Big[-J(c_j^\dag c_{j+1} + \text{h.c.})+V'(t)(n_j-\nu)(n_{j+1}-\nu) \nonumber \\
&+V(t)(n_j-\nu)(n_{j+2}-\nu)\Big]\,,
\label{eqn:extended_tV}
\end{align}
known as extended $t-V$ model. At filling $\nu=1/3$, in thermal equilibrium the simultaneous presence of nearest neighbor $V'$ and next nearest neighbor $V$ interactions stabilizes a charge-density wave (CDW) phase for strong $V'$ and $V$. For our quench study, as in the main text we start from zero interactions (here $V(0) = V'(0) = 0$) and at time $t=0$ the interaction strengths are suddently quenched to a constant finite value $V(t) = V'(t) = V > 0$. The results are shown in Fig.~\ref{figS5}. We observe that the long time behavior of the entanglement shows qualitatively the same behavior as in the case of $V'=0$ studied before, featuring the crossover from dominant real-space to dominant momentum-cut entanglement. These observations are consistent with our intuitive picture, that the competition between kinetic (momentum) and interaction (real-space) terms typical of Hubbard models is also reflected in the entanglement of such systems dynamically approaching equilibrium after a quench.

\begin{figure} [ht!] 
	\includegraphics[width=\linewidth]{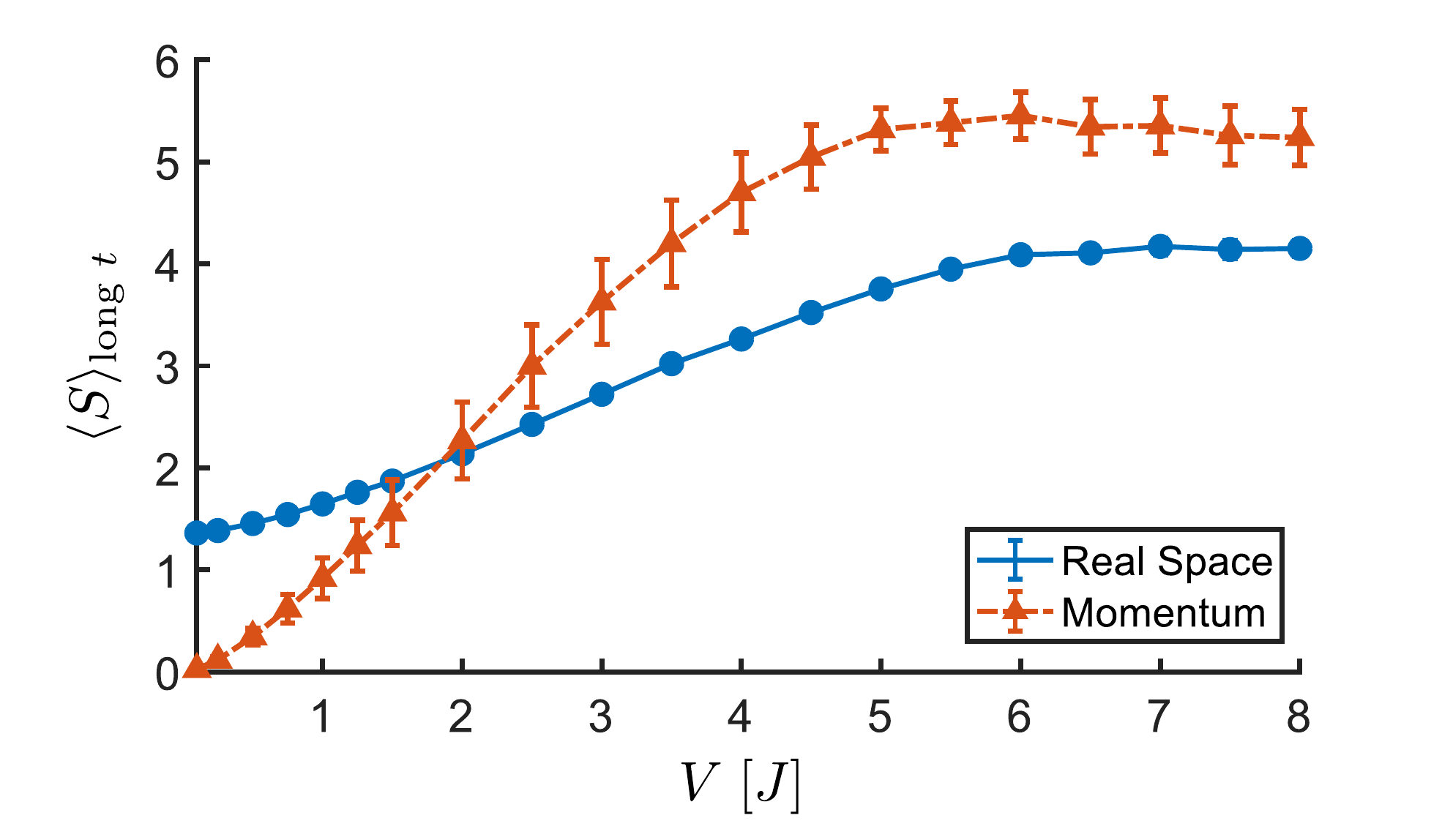}
	\caption{Long time average of $S$ for a interaction quench in the extended $t-V$ model (\ref{eqn:extended_tV}) with $L = 24$ and $\nu = 1/3$ as a function of the post-quench interaction strengths $V'=V$.}
	\label{figS5}
\end{figure}

\bibliographystyle{apsrev}

\begin{thebibliography}{10}
		
		
		\bibitem{WenBook} X.-G.\ Wen, \emph{Quantum Field Theory of Many-Body Systems} (Oxford University Press, Oxford, 2007).
		
		\bibitem{AmicoVedral_EntangRev} L.\ Amico, R.\ Fazio, A.\ Osterloh, V.\ Vedral, Rev. Mod. Phys. \textbf{80}, 517 (2008).
		
		
		\bibitem{NielsenChuangBook} M.\ A.\ Nielsen, I.\ L.\ Chuang, \emph{Quantum Computation and Quantum Information}, (Cambridge University Press, Cambridge, 2000).
		
		\bibitem{BennetEtAl_QuantumTeleportation} C.\ H.\ Bennett, G.\ Brassard, C.\ Cr\'epeau, R.\ Jozsa, A.\ Peres, W.\ K.\ Wootters, Phys. Rev. Lett. \textbf{70}, 1895 (1993).
		
		\bibitem{PlenioVedral_EntQInfo_Rev} M.\ B.\ Plenio, V.\ Vedral, Contemp. Phys. \textbf{39}, 431 (1998).
		
		\bibitem{Horodecki_x_4_QuantumEntanglement} R.\ Horodecki, P.\ Horodecki, M.\ Horodecki, K.\ Horodecki,  Rev. Mod. Phys. \textbf{81}, 865 (2009).
		
		
		\bibitem{MicroscopyEnhancement} T.\ Ono, R.\ Okamoto, S.\ Takeuchi, Nat. Commun. \textbf{4}, 2426 (2013).
		
		
		\bibitem{LiebRobinson} E.\ H.\ Lieb, D.\ W.\ Robinson, Commun. Math. Phys. \textbf{28}, 251 (1972).
		
		\bibitem{CalabreseCardy_EntEvo1D} P.\ Calabrese, J.\ Cardy, J. Stat. Mech. (2005) P04010.
		
		\bibitem{DeChiaraEtAl_EntEvoHeisenberg} G.\ De Chiara, S.\ Montangero, P.\ Calabrese, R.\ Fazio, J. Stat. Mech.: Theor. Exp. (2006) P03001.
		
		\bibitem{LauchliKollath_EntQuench1Dbosons} A.\ M.\ L\"auchli, C.\ Kollath, J. Stat. Mech. (2008) P05018.
		
		\bibitem{KimHuse_BallisticEntang} H.\ Kim, D.\ A.\ Huse, Phys. Rev. Lett. \textbf{111}, 127205 (2013).
		
		\bibitem{Daley_EntGrowthQuench} J.\ Schachenmayer, B.\ P.\ Lanyon, C.\ F.\ Roos, A.\ J.\ Daley, Phys. Rev. X \textbf{3}, 031015 (2013).
		
		\bibitem{Cotler_EntQuenchFieldTheory} J.\ S.\ Cotler, M.\ P.\ Hertzberg, M.\ Mezei, M.\ T.\ Mueller, J. High Energy Phys. \textbf{11} (2016) 166.
		
		\bibitem{Abanin_EntDynamics} W.\ W.\ Ho, D.\ A.\ Abanin, Phys. Rev. B \textbf{95}, 094302 (2017).
		
		\bibitem{LuitzBarLev} D.\ J.\ Luitz, Y.\ Bar Lev, Phys. Rev. B \textbf{96}, 020406(R) (2017).
		
		\bibitem{Mezei_EntSpreadChaotic} M.\ Mezei, D.\ Stanford, J. High Energy Phys. \textbf{05} (2017) 065.
		
		\bibitem{NahumEtAl_EntGrowthRandomUnitary} A.\ Nahum, J.\ Ruhman, S.\ Vijay, J.\ Haah, Phys. Rev. X \textbf{7}, 031016 (2017).
		
		\bibitem{Calzona2017} A. Calzona, F. M. Gambetta, F. Cavaliere, M. Carrega, M. Sassetti, Phys. Rev. B \textbf{96}, 085423 (2017).	
		
		\bibitem{Mistakidis2018} S. I. Mistakidis, G. C. Katsimiga, P. G. Kevrekidis,  P. Schmelcher, New J. Phys. \textbf{20}, 043052 (2018).
		
		\bibitem{PollmannSondhi_PRXall} C.\ W.\ von Keyserlingk, T.\ Rakovszky, F.\ Pollmann, S.\ L.\ Sondhi, Phys. Rev. X \textbf{8}, 021013 (2018).
		
		\bibitem{NahumEtAl_EntDyn1Drand} A.\ Nahum, J.\ Ruhman, D.\ A.\ Huse, Phys. Rev. B \textbf{96}, 035118 (2018).
		
		\bibitem{SuraceTagliacozzo} J. Surace, M. Piani, L. Tagliacozzo, Phys. Rev. B \textbf{99}, 235115 (2019).
		
		
		
		
		\bibitem{WhiteDMRG} S.\ R.\ White, Phys. Rev. Lett. \textbf{69}, 2863 (1992).
		
		\bibitem{StellanMPS} S.\ \"Ostlund, S.\ Rommer, Phys. Rev. Lett. \textbf{75}, 3537 (1995).
		
		\bibitem{HieidaVBS} Y.\ Hieida, K.\ Okunishi, Y.\ Akutsu, New J. Phys. \textbf{1}, 7 (1999).
		
		\bibitem{VerstraeteCitracVBS} F.\ Verstraete, J.\ I.\ Cirac, Phys. Rev. A \textbf{70}, 060302(R) (2004).
		
		\bibitem{VerstraeteCiracPEPS} F.\ Verstraete, J.\ I.\ Cirac, arXiv:cond-mat/0407066.
		
		\bibitem{Daley2004} A. Daley, C. Kollath, U. Schollw\"ock, G. Vidal, J. Stat. Mech. (2004) P04005.
		
		\bibitem{VerstraeteCiracMPS} F.\ Verstraete, J.\ I.\ Cirac, Phys. Rev. B \textbf{73}, 094423 (2006).
		
		\bibitem{ShiDuanVidalTTN} Y.-Y.\ Shi, L.-M.\ Duan, G.\ Vidal, Phys. Rev. A \textbf{74}, 022320 (2006).
		
		\bibitem{VidalMERA1} G. Vidal, Phys. Rev. Lett. \textbf{99}, 220405 (2007).
		
		\bibitem{VerstraeteMurgCirac_revPEPS} F.\ Verstraete, V.\ Murg, and J.\ I.\ Cirac, Adv. Phys. \textbf{57}, 143 (2008).
		
		\bibitem{VidalMERA2} G. Vidal, Phys. Rev. Lett. \textbf{101}, 110501 (2008).
		
		\bibitem{McCulloch_iDMRG} I.\ P.\ McCulloch, arXiv:0804.2509.
		
		\bibitem{TagliacozzoEvenblyVidalTTN} L.\ Tagliacozzo, G.\ Evenbly, G.\ Vidal, Phys. Rev. B \textbf{80}, 235127 (2009)
		
		\bibitem{SchollwoeckMPS} U.\ Schollw\"ock, Ann. Phys. \textbf{326}, 96 (2011).
		
		
		
		
		
		
		
		\bibitem{KrylovTimeProp} A.\ Nauts, R.\ E.\ Wyatt, Phys. Rev. Lett. \textbf{51}, 2238 (1983).
		
		
		\bibitem{som} See the appendix for technical details.		
		
		
		
		
		\bibitem{SSH} W. P. Su, J. R. Schrieffer, A. J. Heeger, Phys. Rev. Lett. {\bf{42}}, 1698 (1979).
		
		\bibitem{SSHReview} A. J. Heeger, S. Kivelson, J. R. Schrieffer, W. P. Su, Rev. Mod. Phys. {\bf{60}}, 781 (1988).
		
		
		
		\bibitem{AdamPaper} D.\ P.\ Kingma, J.\ L.\ Ba, \emph{Proceedings of the 3rd International Conference on Learning Representations (ICLR), San Diego, 2015}, arXiv:1412.6980.
		
		
		\bibitem{AndiPRB} A.\ Kruckenhauser, J.\ C.\ Budich, Phys. Rev. B \textbf{98}, 195124 (2018).
		
		
		\bibitem{NNNphasediag_PRB} P.\ Schmitteckert, R.\ Werner, Phys. Rev. B \textbf{69}, 195115 (2004).
		
		\bibitem{NNNergodic_PRE} C.\ Neuenhahn, F.\ Marquardt, Phys. Rev. E \textbf{85}, 060101(R) (2012).
		
		
		\bibitem{Hastings_AreaLaw1D} M.\ B.\ Hastings, J. Stat. Mech. Theory Exp. (2007) P08024.
		
		\bibitem{Eisert_AreaLawReview} J.\ Eisert, M.\ Cramer, and M.\ B.\ Plenio, Rev. Mod. Phys. \textbf{82}, 277 (2010).
		
		
		\bibitem{Weichselbaum2012}
		C. Guo, A. Weichselbaum, J. von Delft, and M. Vojta,
		Phys. Rev. Lett. {\bf{108}}, 160401 (2012).
		
		\bibitem{Mitrushenkov2001}
		A. O. Mitrushenkov, G. Fano, F. Ortolani, R. Linguerri, and P. Palmieri,
		J. Chem. Phys. {\bf{115}}, 6815 (2001).
		
		
		\bibitem{Szalay2015}
		S. Szalay, M. Pfeffer, V. Murg, G. Barcza, F. Verstraete, R. Schneider, \"O. Legeza,
		Int. J. Quant. Chem. {\bf{115}}, 1342 (2015).
		
		\bibitem{Krumnow2016}
		C. Krumnow, L. Veis, \"O. Legeza, J. Eisert, Phys. Rev. Lett. {\bf{117}}, 190602 (2016). 
		
		
		
		\bibitem{eisert_basisMPS} C.\ Krumnow, J.\ Eisert, \"O.\ Legeza, arXiv:1904.11999.
		
		\bibitem{transport_freqdomain} M.\ M.\ Rams, M.\ Zwolak, arXiv:1904.12793.
		
	\end{thebibliography}

\begin{thebibliography}{4}
	
	\bibitem{SOMNautsWyatt_KrylovTimeProp} A.\ Nauts, R.\ E.\ Wyatt, Phys. Rev. Lett. \textbf{51}, 2238 (1983).
	
	\bibitem{SOMLinGubernatis_ED} H.\ Q.\ Lin, J.\ E.\ Gubernatis, Computers in Physics \textbf{7}, 400 (1993).
	
	\bibitem{SOMAdamPaper} D.\ P.\ Kingma, J.\ L.\ Ba, \emph{Proceedings of the 3rd International Conference on Learning Representations (ICLR), San Diego, 2015}, arXiv:1412.6980.
	
	\bibitem{SOMAndiPRB} A.\ Kruckenhauser, J.\ C.\ Budich, Phys. Rev. B \textbf{98}, 195124 (2018).
	
	\bibitem{SOMPeschel_RedDensMatFree} I.\ Peschel, Journal of Physics A: Mathematical and General \textbf{36}, L205 (2003).
	
	\bibitem{SOMVidalEtAl_EntangCriticalSys} G.\ Vidal, J.\ I.\ Latorre, E.\ Rico, A.\ Kitaev, Phys. Rev. Lett. \textbf{90}, 227902 (2003).
	
	
	
\end{thebibliography}

\end{document}